# Spectral Cross-Cumulants for Multicolor Super-resolved SOFI Imaging

K. S. Grußmayer[1,2, #, *], S. Geissbuehler[2, #], A. Descloux [1,2], T. Lukes[1,2], M. Leutenegger[2,3], A. Radenovic[1], T. Lasser[2,4,*]

Affiliations

[1]École Polytechnique Fédérale de Lausanne, Laboratory of Nanoscale Biology, 1015 Lausanne, Switzerland

[2]École Polytechnique Fédérale de Lausanne, Laboratoire d'Optique Biomédicale, 1015 Lausanne, Switzerland

[3]Max-Planck Institute for Biophysical Chemistry, Department of NanoBiophotonics, Am Fassberg 11, 37077 Göttingen, Germany

[4]Max-Planck Institute for Polymer Research, Ackermannweg 10, 55128 Mainz, Germany

Contributions

[#]These authors contributed equally to this work.

Corresponding Authors:

Kristin Grußmayer, Theo Lasser

* email: kristin.grussmayer@epfl.ch, theo.lasser@epfl.ch

*Abstract*

Super-resolution optical fluctuation imaging (SOFI) provides a resolution beyond the diffraction limit by analysing stochastic fluorescence fluctuations with higher-order statistics. Using $n^{\text{th}}$ order spatio-temporal cross-cumulants the spatial resolution as well as the sampling can be increased up to *n*-fold in all three spatial dimensions. In this study, we extend the cumulant analysis into the spectral domain and propose a novel multicolor super-resolution scheme. The simultaneous acquisition of two spectral channels followed by spectral cross-cumulant analysis and unmixing increase the spectral sampling. The number of discriminable fluorophore species is thus not limited to the number of physical detection channels. Using two color channels, we demonstrate spectral unmixing of three fluorophore species in simulations and multiple experiments with different cellular structures, fluorophores and filter sets. Based on an eigenvalue/ vector analysis we propose a scheme for an optimized spectral filter choice. Overall, our methodology provides a novel route for easy-to-implement multicolor sub-diffraction imaging using standard microscopes while conserving the spatial super-resolution property. This makes simultaneous multiplexed super-resolution fluorescence imaging widely accessible to the life science community interested to probe colocalization between two or more molecular species.

*Introduction*

Multicolor fluorescence microscopy is an invaluable tool for the study of cellular structures and function. Classical optical microscopes provide single-fluorophore sensitivity, however the spatial resolution is limited due to diffraction[1]. During the last two decades, several super-resolution microscopy techniques overcame this limitation by exploiting on- and off-switching of fluorophores in an either deterministic or stochastic manner[2,3]. These methods often require fluorophores with high photostability (e.g. in stimulated emission depletion microscopy (STED)) or high off-to-on-switching ratios (e.g. in single-molecule localization microscopy (SMLM)). Super-resolution microscopes are only slowly finding their way into routine biological application due to complexity in instrument and sample preparation. Overcoming these hurdles with novel schemes may increase the adoption of advanced microscopy techniques. The above mentioned restrictions pose important road blocks for multicolor imaging applications. Several sub-diffraction imaging methods have demonstrated multicolor imaging[4,5]. They mainly rely on fluorophores with distinct spectra[6-9], more complex probes and labels for multiplexing[10-12] which are recorded sequentially in multiple channels, or on more complex approaches taking advantage of other fluorophore properties such as fluorescence lifetime[13].

The difficulty in obtaining optimal fluorophore behavior across the spectrum compatible with the constraints imposed by SMLM limits multicolor camera-based nanoscopy at present, i.e. it is problematic to identify suitable fluorophore multiplets. Workarounds such as spectrally resolved STORM[14] allow the use of several far-red-emitting fluorophores, albeit at the cost of much-increased hardware and analysis complexity. Thus, there is a need for robust and easy-to-implement multicolor sub-diffraction imaging.

Super-resolution optical fluctuation imaging (SOFI)[15,16] provides an elegant way of overcoming the diffraction limit in all spatial dimensions[17]. A classical widefield fluorescence microscope is used to acquire an image sequence of stochastically blinking fluorophores. Post-processing by calculating higher-order cumulants leads to a resolution improvement growing with the cumulant order. Unlike single-molecule localization-based techniques, SOFI does not require the spatiotemporal isolation of individual fluorophores' emissions[18,19] and is thus compatible with a wider range of blinking conditions and labeling densities. Thereby, SOFI simplifies the fluorophore selection, which is particularly welcome due to the inherent difficulty in labeling more than one protein in sufficient quality for super-resolution microscopy. Furthermore, due to the inherent optical sectioning properties of SOFI, the imaging of thick samples can be performed with widefield illumination and does not rely on physical background reduction, such as total internal reflection[20].

To date, cumulant analysis has been used for spatial super-resolution[15]. In this work we generalize the cumulant analysis by extending it into the spectral domain to pave the way towards a novel multicolor SOFI. Unlike for other multicolour approaches, the crosstalk between the different physical color channels is a key ingredient for generating additional virtual color channels.

To our knowledge, only two-color SOFI has so far been demonstrated and used to visualize different structures in a cell and most experiments have been conducted sequentially[19,21-23]. However, by imaging multiple spectral channels step by step, one key feature of cumulants is not exploited, i.e. the cross-cumulation in between detection channels.

In spatiotemporal cross-cumulation[16] various combinations of cross-cumulants between neighbouring pixels are used to obtain virtual pixels (i.e. a denser sampling of the super-resolved image). Here, we apply cross-cumulants between multiple simultaneously acquired spectral channels. The physical detection channels are thereby supplemented by virtual spectral channels to obtain a finer spectral sampling. The refined spectral sampling allows linear unmixing[24,25] of many distinct fluorophore colors $N_C$ with at least 2 recorded physical acquisition channels. For an $n^{th}$-order cumulant analysis of $N_p$ physical channels a total of $N_c$ cumulant color channels can be generated, with

$$N_\text{c} = \prod_{i=2}^{N_\text{p}} \frac{n+i-1}{i-1} \tag{1}$$

the number of distinct *n*-tuple combinations of the $N_\text{p}$ physical channels without permutations.

## Results

*Spectral unmixing using spectral cross-cumulants*

Classical simultaneous multicolor imaging is achieved by adding dichroic filter in the imaging path to separate the light into two or more distinct spectral channels. For the following discussion, we will consider the simplest case where only two physical channels are used (see Supporting Information Section Theory of spectral unmixing using spectral cross-cumulants for a general discussion). For a given dichroic leading to an overall transmission spectrum $D_T(\lambda)$ and fluorophore species with an emission spectrum $S_i(\lambda)$, we can define the transmission $T_i = \frac{\int_0^\infty S_i(\lambda) D_T(\lambda) d\lambda}{\int_0^\infty S_i(\lambda) d\lambda}$ and reflection coefficient $R_i = \frac{\int_0^\infty S_i(\lambda)(1-D_T)(\lambda) d\lambda}{\int_0^\infty S_i(\lambda) d\lambda}$ describing the collection of fluorescence signal which is detected according to the spectral response of the respective channel (determination of $T_i$ and $R_i$ see Supporting Information Section Transmission and reflection coefficients). If absorption and scattering of the dichroic is neglected, the relation $T_i = 1 - R_i$ holds. If we consider three spectrally distinct fluorophore species to be detected by two spectral detection channels, we can express the intensities $I_{R;T}(\boldsymbol{r})$ recorded in both spectral channels as

$$\begin{aligned} I_R(\boldsymbol{r}) &= \sum_{i=1}^{3} R_i I_i(\boldsymbol{r}) \\ I_T(\boldsymbol{r}) &= \sum_{i=1}^{3} T_i I_i(\boldsymbol{r}) \end{aligned} \tag{1}$$

with $I_i(\boldsymbol{r})$ being the intensity distribution of dye species *i* measured at the detector pixel *r*. This linear system can only be solved when using additional information to retrieve the images of the distinct dye species[24], as there are three unknowns $I_1(\boldsymbol{r})$, $I_2(\boldsymbol{r})$ and $I_3(\boldsymbol{r})$ for only two measurements $I_R(\boldsymbol{r})$ and $I_T(\boldsymbol{r})$.

Assuming stochastic, independent blinking of all fluorescent emitters, we can apply cumulant analysis on the time series $I_{R;T}(\boldsymbol{r},t)$ recorded in the two physical channels and generate a so-called virtual spectral channel by computing the second-order cross-cumulants (see Figure 1). This virtual channel contains only cross-talk contributions from emitters that are detected and most important are correlated in both physical channels. Due to the additivity[15], the cumulant of multiple independent fluorophore species corresponds to the sum of the cumulants of each individual species and we can rewrite:

$$\begin{pmatrix} \kappa_{2,RR}(\boldsymbol{r}) \\ \kappa_{2,TR}(\boldsymbol{r}) \\ \kappa_{2,TT}(\boldsymbol{r}) \end{pmatrix} = \begin{pmatrix} R_1^2 & R_2^2 & R_3^2 \\ T_1 R_1 & T_2 R_2 & T_3 R_3 \\ T_1^2 & T_2^2 & T_3^2 \end{pmatrix} \begin{pmatrix} \kappa_2\{I_1(\boldsymbol{r},t)\} \\ \kappa_2\{I_2(\boldsymbol{r},t)\} \\ \kappa_2\{I_3(\boldsymbol{r},t)\} \end{pmatrix} \qquad (2)$$

where $\kappa_{2,RR}$ and $\kappa_{2,TT}$ are second order SOFI images calculated using intensities of two-pixel combinations from the reflection and transmission channel[16], respectively. $\kappa_{2,TR}$ is the second order spectral cross-cumulant image calculated using one pixel from each physical color channel. $\kappa_2\{I_i(\boldsymbol{r},t)\} = \kappa_{2;i}$ denotes the second order cumulant of the different dye species i. The possibility to compute an additional channel is the key to enabling unmixing by inversion of the linear system of equations (see Equation 2). We can thus recover the individual second order cumulants for the three fluorophore species provided the spectral sensitivities of the dyes allow inversion of the matrix in equation (2). We provide a guide for selecting the optimal combination of dyes and filter set based on an eigenvalue/ vector analysis in the Supporting Information ( see Section Optimization of multicolor SOFI via eigenvalue and eigenvector analysis)) So far, we have considered the simplest case of two physical channels, however, this novel cross-spectral cumulant analysis can be in principle generalized to obtain additional virtual spectral channels as shown in the Supporting Information Section Theory of spectral unmixing using spectral cross-cumulants.

*Workflow spectral cross-cumulant analysis for multicolor SOFI*

In our experiments, we implemented two physical detection channels using a dichroic filter that spectrally splits the fluorescence emission and dispatches the reflected and transmitted light on two synchronized sCMOS cameras (see Figure 1a). To demonstrate the feasibility of the described multicolor SOFI with spectral unmixing, we first generated simulated datasets (see Materials and Methods).

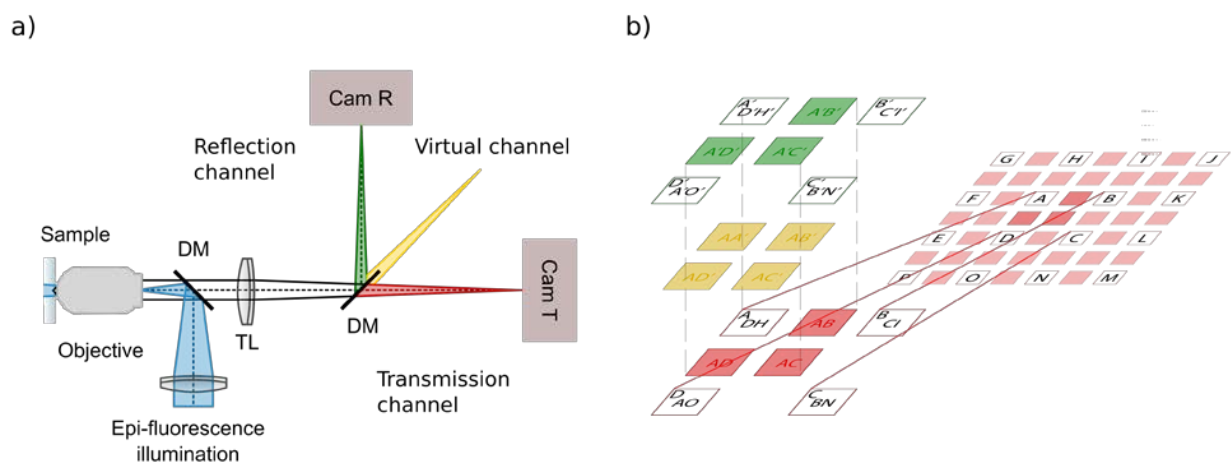

*Figure 1 Cross-cumulant analysis between spectral channels. (a) Simplified detection scheme with two physical spectral channels directed on two separate cameras (Cam R and Cam T, green and red). Spectral cross-cumulant analysis allows the generation of additional virtual spectral channels (yellow). DM dichroic mirror, TL tube lens, R reflection, T transmission. (b) Pixel combinations for the second-order cross-cumulant calculation. The cumulant analysis of each spectral channel (spectral auto-cumulant) is performed as described previously[16]. By cross-correlating intensities from different spectral channels (red and green) analogous to the computation of 'virtual' planes in multi-plane 3D SOFI[17], the additional 'virtual' cross-cumulant channel is computed (yellow). Single letters denote the original pixel matrix whereas multiplets of letters symbolize cross-cumulant calculation from combinations of original pixel intensities.*

Figure S2 provides an example of the three common organic fluorophores Alexa Fluor 488, Atto565 and Alexa Fluor 647. Their emission spectra are weighted with the spectral responses of the reflection and transmission channels obtained by a multi-band dichroic and emission filter and a dichroic mirror at $\lambda = 594$ nm. We generated image sequences (reflection and a transmission channel) for patches arranged in a grid of randomly blinking fluorophores representing either one of the three spectra,. The workflow of the multicolor SOFI analysis is outlined in Figure 2. After data acquisition, co-registration of the physical color channels is performed (Figure 2 step 1, for details about the co-registration see Supporting Information Section Co-registration of physical color channels). The individual patches cannot be distinguished in the widefield images.

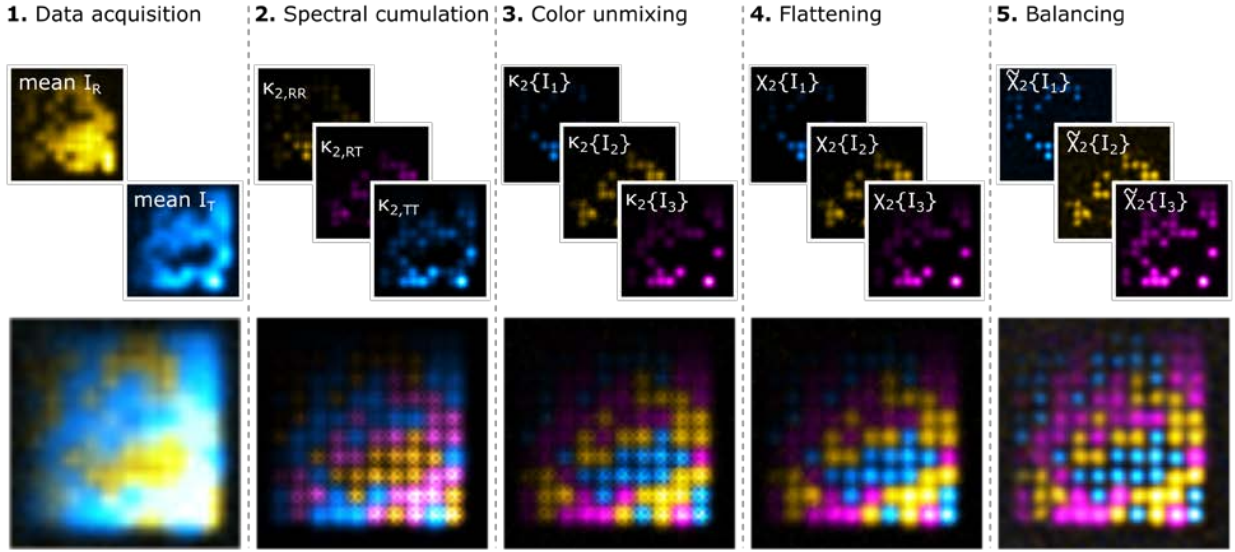

Figure 2 Workflow of multicolor SOFI imaging by spectral cross-cumulant analysis followed by linear unmixing using simulations. A 2.5 pixel grid of patches with ~20nm radius and 2 fluorophores each (~1300 emitter µm$^{-2}$) was simulated. Alexa488 (blue hot), Atto565 (yellow hot) or Alexa647 (magenta hot) spectral properties are randomly assigned to each patch. $I_{on}$ varies from top to bottom (200 to 1100 photons) and the on-ratio varies from left to right (0.01 to 0.1). 4000 frames with negligable photobleaching were analyzed.

In the second step, the second-order spectral cross-cumulants are calculated to generate the three spectral channels $\kappa_{2,RR}(r)$, $\kappa_{2,TR}(r)$ and $\kappa_{2,TT}(r)$ as described above. The second-order cross-cumulant analysis adds a virtual color channel which already leads to a better spatial resolution inherent to raw SOFI images[15]. The individual patches can be identified, but the dyes are not yet unmixed. If the spectral cross-cumulants are computed with different spatial shifts and/or temporal delays, the resulting image generally has an inhomogeneous weight distribution arising from the spatio-temporal decorrelation of the signal[16] (as can be seen in Figure 2, steps 2-3). We use zero time-delays in order to be more flexible in the range of detectable temporal fluctuations, but use spatial cross-cumulants to increase the virtual pixel grid density[16]. It is important to note, that the different pixels involved in the spectral cross-cumulants in step 2 of the algorithm should have the same spatial shifts for a specific output pixel such that the matrix inversion for color unmixing following in step 3 (see Figure 2, step 3) is possible. The fourth step consists in correcting for these inhomogeneous pixel weights by applying a distance-factor correction individually for each color or by maintaining the same mean for all pixel sub-grids, as it is applied for single-color spatial cross-cumulant SOFI. This flattening procedure cancels the spatial decorrelation arising from the finite PSF size[16] (see Figure 2 step 4 and Supporting Information for details). The last step (see Figure 2 step 5) consists in linearizing the cumulant response to brightness[25] by deconvolving each separate dye species image using Lucy-Richardson deconvolution, then taking the $n^{th}$ root and finally by reconvolving with a physically reasonable PSF. The patches can now be distinguished according to the spectra of the fluorophores and the patch size correlates with the wavelength (blue<yellow<magenta). Our analysis shows that spectral cross-cumulant analysis followed by unmixing tolerates a large range of intensities and blinking behaviour. We can thus generate multicolor images with all the advantages inherent to SOFI such as optical sectioning, elimination of uncorrelated background and increased spatial resolution[15]. We investigated the influence of different photophysical properties of the fluorophores such as very closely overlapping spectra, varying photostability, on-time ratio and brightness on the performance of our multicolor analysis (see Supporting Information Section Simulations). For commonly used triplets such as Alexa Fluor 488, Atto 565 and Alexa Fluor 647 the crosstalk remaining after our novel analysis is very low with 5% or less (see Table S2). When the fluorophore emission maxima are only ~10 nm apart, e.g. for Abberior Flip 565 ($\lambda_{\frac{abs}{em},max} = 566/580$ nm), Atto 565 ($\lambda_{\frac{abs}{em},max} = 564/590$ nm) and Alexa Fluor 568 ($\lambda_{\frac{abs}{em},max} = 578/603$ nm), the remaining crosstalk is at worst 14% (see Table S3). Our exemplary analysis of filament-like structures shows that our multicolor processing is reliable for a wide range of expected fluorophore behaviour.

*Experimental results*

As a first experimental demonstration of multicolor SOFI with spectral unmixing, we imaged a time series of fixed HeLa cells stained with three different fluorophores. We collected the fluorescence light splitted across a dichroic at $\lambda = 594$ nm on two synchronized sCMOS cameras as mentioned above (see Figure 3 a). The spectral response of the reflection and transmission channel as well as the weighted emission of the three fluorophores matches the values of the simulations shown above (see Figures 2 and S1). We labeled microtubules with Alexa Fluor 488 via antibody staining, glycoproteins and sialic acid using wheat germ agglutinin (WGA)-Atto 565 and Lamin B1 in the nuclear membrane with Alexa Fluor 647 also via antibody staining. Appropriate blinking for SOFI processing was achieved using a buffer with thiols and oxygen scavengers; the fluorophores were excited with three different lasers at 488 nm, 561 nm and 635 nm wavelength and moderate illumination intensities. We evaluated 2000 frames acquired with an exposure time of 20 ms per frame. The cumulant calculation was split into 10 subsequences to minimize the impact of photobleaching on SOFI analysis. Based on the conventional dual channel image (Figure 3a), separation of the different labels is impossible. Spectral cross-correlation leads to the known optical sectioning and background reduction inherent to SOFI analysis[15] and generates the third virtual channel $\kappa_{2,RT}$ that is dominated by the WGA signal (Figure 3b, blue and Figure S11). Figure S11 illustrates the remaining spectral crosstalk present in the three channels. The final unmixed color channels (see Figure 3c-f) manage to separate the cell's microtubule network (Alexa Fluor 488, blue) from the diffuse WGA-staining (Atto 565, green) and the typical Lamin B1 structure in the nuclear membrane (Alexa Fluor 647, red). WGA is a lectin that labels the cell membrane, but also the Golgi apparatus in the periphery of the nucleus and nuclear pore complex proteins in the nuclear membrane[26]. Residual crosstalk and imperfectly reconstructed microtubules are only apparent in a small part of the image in the periphery of the nucleus (see arrow in Figure 3f), where many microtubules overlap with extremely bright Golgi staining. Otherwise, the algorithm performs remarkably well although the image contains overlapping structures across the field of view and we use common fluorophores that have a high on-ratio under the applied moderate illumination conditions (Atto 565 and Alexa Fluor 488).

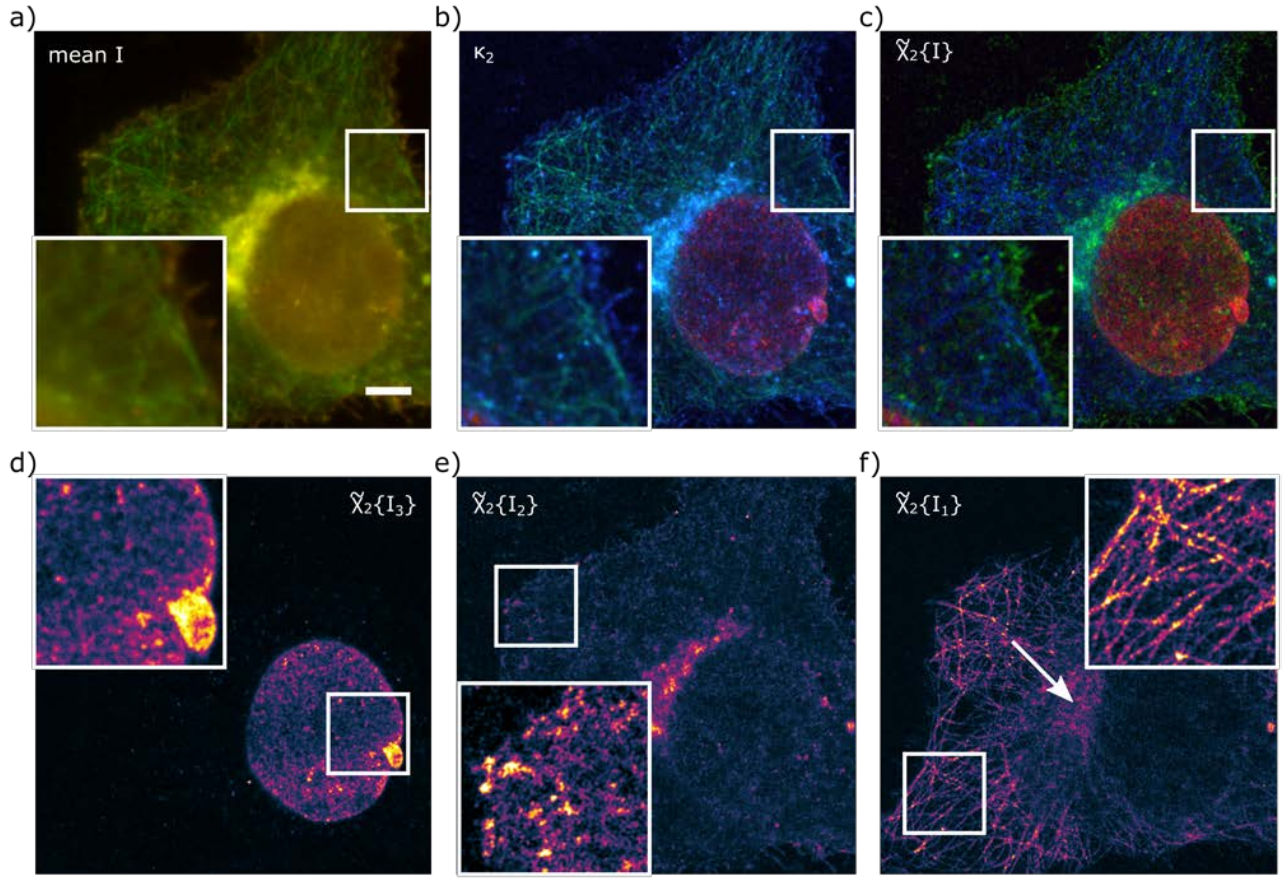

*Figure 3 Multicolor SOFI of the cytoskeleton, nucleus and cellular membranes of HeLa cells. a) Overlay of the average intensity acquired in the reflection (green) and transmission (red) channel using 200 mM MEA with oxygen scavenging and about 0.5 kWcm$^{-2}$ 488 nm , 1.25 kWcm$^{-2}$ 561 nm and 1.3 kWcm$^{-2}$ 635 nm illumination intensity. b) RGB composite image of the second order spectral cross-cumulant images with $\kappa_{2,RR}$(green), $\kappa_{2,RT}$(blue) and $\kappa_{2,TT}$(red) c) RGB composite image of the unmixed, flattened and deconvolved SOFI images with d) Alexa Fluor 647 secondary antibody stained nuclear membrane (red, $\tilde{\chi}_2\{I_3\}$), e) wheat germ agglutinin-Atto565 labeling (green, $\tilde{\chi}_2\{I_2\}$) and f) Alexa Fluor 488 secondary antibody stained microtubules (blue, $\tilde{\chi}_2\{I_1\}$). The separate unmixed images are displayed using the morgenstemning colormap[27]. Scale bar 5 μm.*

As a second example, we included two densely labelled, almost completely overlapping structures in the nucleus of the cell. The DNA was stained with Hoechst-Janelia Fluor 549 with a blue-shifted spectrum compared to the previously used Atto 565 and the nuclear membrane was visualized as before (antibody staining of Lamin B1 with Alexa Fluor 647), with the focus on the bottom of the nucleus. The third structure is again microtubules that are labeled with Alexa Fluor 488 via antibodies. One can appreciate that our spectral cross-cumulant unmixing approach manages to disentangle the two nuclear stains due to the specific spatiotemporal fluctuations of the different fluorophores. Distinctly different small structures are revealed showing the expected patterns such as small folds in the nuclear membrane (see Figure 4 and Figure S12), whereas the overall shape appears in both the Janelia Fluor 549 and the Alexa Fluor 647 channel.

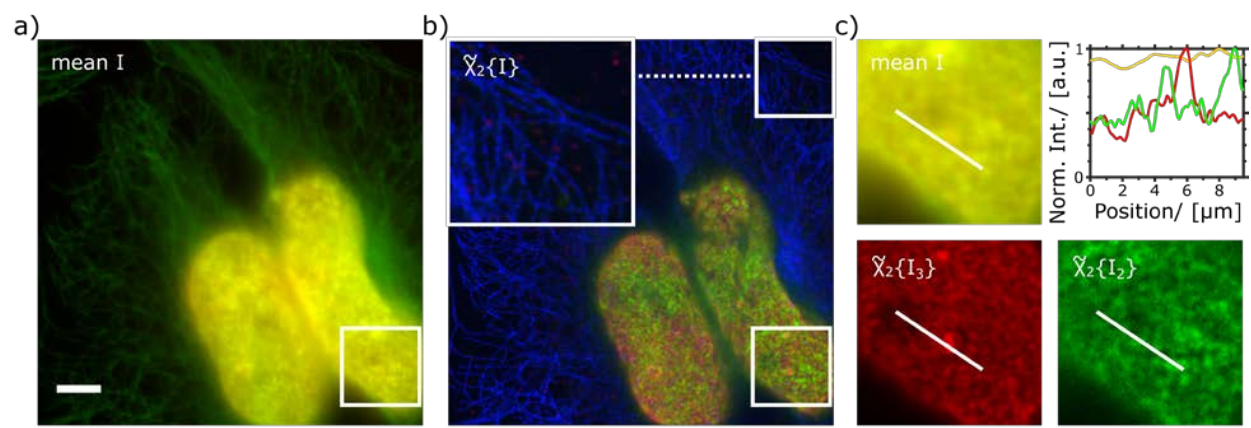

*Figure 4 Multicolor SOFI of the cytoskeleton and nucleus of HeLa cells. a) Overlay of the average of the time series acquired in the reflection (green) and transmission (red) channel using 50 mM MEA with oxygen scavenging and about 0.5 kWcm$^{-2}$ 488 nm , 1.25 kWcm$^{-2}$ 561 nm and 1.3 kWcm$^{-2}$ 635 nm illumination intensity. b) RGB composite image of the unmixed, flattened and deconvolved second order SOFI images with Alexa Fluor 488 secondary antibody stained microtubules (blue, $\tilde{\chi}_2\{I_1\}$), Hoechst-Janelia Fluor 549 DNA labeling (green, $\tilde{\chi}_2\{I_2\}$) and Alexa Fluor 647 secondary antibody stained nuclear membrane (red, $\tilde{\chi}_2\{I_3\}$). c) Close-up of the ROI indicated in a) and b), respectively and comparison of the normalized intensity profiles along the indicated line. Scale bar 5 µm.*

Additional experiments confirm that spectral cross-cumulant analysis can be applied using different dichroic beam splitters (see Figure S13 and S14) and for fluorophores having a larger spectral overlap (see Figure S15 and S16). Only two lasers were needed for the excitation of the three species (Alexa Fluor 568 secondary antibody staining of mitochondria, Hoechst-Janelia Fluor 549 labeling and Alexa Fluor 647 microtubule staining), further reducing the complexity of the experimental setup.

*Summary and Discussion*

In this work, we have shown that the spectral sampling can be refined based on a spectral cross-cumulant calculation between simultaneously acquired color channels. Besides a reduced acquisition time compared to sequential imaging, the simultaneous acquisition of multiple color channels and spectral cross-cumulation allows to unmix several fluorophore species even with strongly overlapping emission spectra, where the number of species is not limited to the number of physical spectral channels. Using a basic two-color detection scheme we validate the spectral unmixing of three fluorophore species labeling in simulations and experiments. We demonstrate multicolor imaging of different structures, fluorophores and filter sets in HeLa and COS-7 cells. We also provide a guideline for optimized spectral filter choice based on an eigenvalue/ vector analysis. The proposed strategy corresponds to the experimentally most straightforward implementation, nowadays frequently met in commercial setups via image splitting units using one camera. The possibility of cross-cumulating between color channels thus translates the concept of spatial super-resolution to spectral super-sampling. Our analysis preserves all the advantages inherent to SOFI such as optical sectioning, elimination of uncorrelated background and increased spatial resolution[15]. Since we formulated multicolor spectral cross-cumulant SOFI in the theoretical framework originally devised for spatially super-resolved SOFI, it is intrinsically compatible with previous developments such as 3D SOFI[17] and bSOFI[25]. In particular, our approach could be combined with the recently published multiple-tau (mt)-pcSOFI[28] to double the number of imaged structures, provided suitable pairs of fluorophores (organic dyes, fluorescent proteins, polymer dots, etc) are available. mt-pcSOFI achieves multiplexing using a single color channel by exploiting differences in blinking kinetics of dyes when calculating SOFI cumulants.

To conclude, we presented a generalized cumulant concept for multicolor spectral cross-cumulant SOFI analysis with thorough proof-of principle simulations and experiments. This novel approach proved to be robust and compatible with a large range of fluorophores enabling sub-diffraction imaging of several cellular structures while using readily available microscope hardware.

## Materials and methods

*Simulations*

The code for simulations was based on the previously published "SOFI Simulation Tool" software package that simulated images of fluorophores from a single species recorded on one camera[29]. Briefly, for each fluorophore species, single emitters are randomly placed according to a certain spatial density and spatial distributions. For each emitter, the blinking behaviour is modelled as a time-continuous Markovian process with exponential probability distribution functions with average blinking on-time $\tau_{on}$ and off-time $\tau_{off}$. The camera detects on average $I_{on}$ photons in the on-state. Fluorophore photobleaching is also considered by a single exponential decay with average bleaching time $\tau_{pb}$. The PSF is assumed to be a rotationally symmetric 2D Gaussian and a standard deviation according to the numerical aperture, camera pixel size and wavelength of the fluorophore. A spatially constant background $I_b$ is added to the total fluorescence signal. The summed signal per frame is then split in the transmission and reflection channel according to the emission spectra of the fluorophore and the spectral response curve of the microscope. The generated image stacks $I_{R,i}(\boldsymbol{r}, t)$ and $I_{T,i}(\boldsymbol{r}, t)$ for the different fluorophore species *i* are then summed up and each pixel intensity is subjected to Poissonian noise. The intensity per pixel is converted to electric charge according to the quantum efficiency and gain of the camera and Gaussian noise with a standard deviation related to dark noise is added to obtain the final time series $I_R(\boldsymbol{r}, t)$ and $I_T(\boldsymbol{r}, t)$.

*Chemicals*

Unless noted otherwise, all chemicals were purchased at Sigma-Aldrich.

*Cell culture*

HeLa cells and COS-7 cells were cultured at 37 °C and 5 % $CO_2$ using DMEM high glucose with pyruvate (4.5 g $l^{-1}$ glucose, with GlutaMAX$^{TM}$ supplement) supplemented with 10 % fetal bovine serum and 1× penicillin-streptomycin (all gibco®, Thermo Fisher Scientific) or DMEM high glucose w/o phenol red (4.5 g $l^{-1}$ glucose) supplemented with 4 mM L-gluthamine, 10 % fetal bovine serum and 1× penicillin-streptomycin (all gibco®, Thermo Fisher Scientific).

Cells were seeded in Lab-tek® II chambered cover slides (nunc) 1-2 days before fixation in DMEM high glucose w/o phenol red (4.5 g $l^{-1}$ glucose) supplemented with 4 mM L-gluthamine, 10 % fetal bovine serum and 1× penicillin-streptomycin (all gibco®, Thermo Fisher Scientific).

*HeLa cell fixation and staining*

HeLa cells were washed twice in pre-warmed buffer (microtubule stabilizing buffer (MTSB): 100 mM PIPES pH 6.8, 2 mM $MgCl_2$, 5 mM EGTA or PBS for wheat germ agglutinin (WGA) staining), followed by application of pre-warmed fixation buffer (3.7 % paraformaldehyde (PFA), 0.2 % Triton X-100 in MTSB or 3.7 % paraformaldehyde (PFA) in PBS for wheat germ agglutinin (WGA) staining) for 15 min at room temperature (RT). Cells were then washed three times for 5 min each with 1× PBS and stored in 50 % glycerol in 1× PBS at 4 °C or the immunostaining protocol was continued to prepare samples for fluorescence imaging.

Fixed and permeabilized cells were blocked with 3 % BSA in 1× PBS and 0.05 % Triton X-100 for 60 min at RT or overnight at 4 °C.

WGA staining: Cells that were fixed without permeabilization were stained with 5 ng $ml^{-1}$ WGA-Atto565 (preparation see below) for 10 min followed by three times 5 min washes with 1× PBS. Subsequently, the cells were blocked using blocking buffer containing 0.2 % Triton X-100.

Microtubule and nuclear envelope staining: The blocked samples were immediately incubated with a mix of primary anti-tubulin antibody (1 mg ml$^{-1}$ DM1a mouse monoclonal (ab80779) 1:150 dilution, Abcam) and anti-Lamin B1 antibody (1 mg ml$^{-1}$ rabbit polyclonal (ab16048) 1:400 dilution, Abcam) in antibody incubation buffer for 60 min at RT (AIB: 1 % BSA in 1× PBS and 0.05 % Triton X-100). Cells were then washed three times for 5 min each with AIB, followed by incubation with a mix of donkey anti-mouse-Alexa Fluor 647 antibody (0.005 mg ml$^{-1}$ Invitrogen) and donkey anti-rabbit-Alexa Fluor 647 antibody (0.01 mg ml$^{-1}$ Invitrogen) for 60 min at RT. This and all subsequent steps were performed in the dark. Cells were again washed three times for 5 min each with AIB, optionally subjected to DNA staining and incubated for 15 min post-fixation with 2 % PFA in 1× PBS followed by three 5 min washes with PBS. Cells were imaged immediately or stored in 50 % glycerol in 1× PBS at 4 °C until imaging.

DNA staining: 10µM Hoechst-Janelia Fluor 549 in PBS was incubated for 10 min followed by three times 5 min washes with 1× PBS.

*COS-7 cells fixation and staining:*

The protocol is similar as described previously by Chazeau et al.[30]. Cells were washed twice in pre-warmed DMEM w/o phenol red (see cell culture) following 90 s incubation with extraction buffer (microtubule stabilizing buffer 2 (MTSB2: 80 mM PIPES, 7 mM MgCl2, 1 mM EGTA, 150 mM NaCl, 5 mM D-glucose adjust pH to 6.8 using KOH) with freshly added 0.3 % Triton X-100 (AppliChem) and 0.25 % glutaraldehyde (stock solution 50 % electron microscopy grade, Electron Microscopy Sciences). Immediately afterwards, pre-warmed 4 % paraformaldehyde (PFA) in PBS was incubated for 10 min at room temperature (RT). Cells were then washed three times for 5 min each with 1× PBS and stored in 50 % glycerol in 1× PBS at 4 °C or the immunostaining protocol was continued. Next, a freshly prepared solution of 10mM NaBH4 in 1× PBS was incubated on the cells for 7 minutes followed by one quick wash in 1× PBS, and two washes 10 min 1× PBS on an orbital shaker. Cells were permeabilized in PBS with 0.25 % Triton X-100 for 7 min followed by blocking with blocking buffer (BB: 2 % (w/v) BSA, 10 mM glycine, 50 mM ammonium chloride $NH_4Cl$ in PBS pH 7.4 for 60 min at RT or overnight at 4 °C.

The blocked samples were incubated with primary anti-tubulin antibody (clone B-5-1-2 ascites fluid 1:200 dilution, Sigma-Aldrich) and primary anti-TOMM20 antibody ([EPR15581], 1:50 dilution, Abcam) in BB for 60 min at RT. Cells were then washed three times for 5 min each with BB, followed by incubation with donkey anti-mouse-Alexa Fluor 647 antibody (donkey anti-mouse (H+L) highly cross-adsorbed at 0.005 mg $ml^{-1}$ Invitrogen) and donkey anti-rabbit-Alexa Fluor 568 (donkey anti-mouse (H+L) highly cross-adsorbed at 0.005 mg $ml^{-1}$ Invitrogen) for 60 min at RT. This and all subsequent steps were performed in the dark. Cells were again washed three times for 5 min each with BB and incubated for 10 min post-fixation with 2 % PFA in 1× PBS followed by three 5 min washes with PBS. Cells were imaged immediately or stored in 50 % glycerol in 1× PBS at 4 °C until SOFI imaging. Just before imaging, 2 µM Hoechst-Janelia Fluor 549 in PBS was incubated for 10 min followed by three times 5 min washes with 1× PBS.

*Preparation of labeled proteins*

2 mg $ml^{-1}$ WGA (Vector Labs) was incubated for 1 h at RT while shaking with Atto565-NHS esther (Atto-tec) at a ratio of 1: 6 with the pH raised to 8.3 using sodium bicarbonate. The mixture was purified using illustra NAP Columns (GE Healthcare) according to manufacturer's instructions and eluted with slightly acidic PBS to recover labeled antibody at neutral pH. The protein concentration was estimated by absorption spectrometry to 0.5 mg $ml^{-1}$ WGA-Atto565.

*Imaging buffer*

The samples were imagined in a 50 mM Tris-Hcl pH 8.0, 10 mM NaCl buffer containing an enzymatic oxygen scavenging system (2.5 mM protocatechuic acid (PCA) and 50 nM Protocatechuate-3,4-Dioxygenase from Pseudomonas Sp. (PCD) with >3 Units m g$^{-1}$) and a thiol (2-Mercaptoethylamine). The thiol and a stock solution of 100 mM PCA in water, pH adjusted to 9.0 with NaOH, were always prepared fresh. PCD was aliquoted at a concentration of 10 µM in storage buffer (100 mM Tris-HCl pH 8.0, 50 % glycerol, 50 mM KCl, 1 mM EDTA) at -20 °C and thawn immediately before use.

*Microscope setup*

All imaging was performed with a custom built wide-field fluorescence microscope equipped with a 200 mW 405 nm laser (MLL-III-405-200mW), a 1 W 635 nm laser (SD-635-HS-1W, both Roithner Lasertechnik), a 350 mW 561 nm laser (Gem561, Laser Quantum) and a 200mW 488nm laser (iBEAM-SMART-488-S-HP, Toptica Photonics). The lasers were combined and focused into the back focal plane of a Nikon SR Plan Apo IR 60× 1.27 NA WI objective. The fluorescence light was filtered using a combination of a dichroic mirror and a multiband emission filter (Quad Line Beamsplitter R405/488/561/635 flat and Quad Line Laser Rejectionband ZET405/488/561/640, both AHF Analysetechnik).

In the detection path, the light is focused by a 200nm tube lens before being split by an emission dichroic (Laser Beamsplitter zt 594 RDC or Beamsplitter HC BS 640 imaging, both AHF Analysentechnik) and directed on two synchronized sCMOS cameras (ORCA Flash 4.0, Hamamatsu; back projected pixel size of 108 nm). For translating the sample, the microscope is equipped with an xy motorized stage (SCAN$^{plus}$ IM 120 × 80 Maerzhaeuser with Tango Desktop driver). Focus stabilization is provided by a nanometer z positioning stage (Nano-ZL300-M; Mad City Labs with Nano-Drive C controller) driven by an optical feedback system similar to[31].

*Data processing*

The algorithm was implemented in MATLAB (Mathworks). We adapted and extended the bSOFI MATLAB package used in[25].

*Author contributions*

K.S.G. performed experiments and simulations. S.G. and M.L. established the multicolor spectral cross-cumulant concept and wrote the initial analysis software. A.D., T.Lu and K.S.G. adapted and extended the analysis software and A.D. and K.S.G. analysed the data. T.L. contributed to the theory of the multicolor spectral cross-cumulant concept and formulated the eigenvalue/eigenvector analysis with A.D. and K.G. . T.L. and A.R. supervised the project. K.S.G wrote the manuscript with input from all authors.

*Acknowledgements*

Hoechst-Janelia Fluor 549 was a kind gift from Luke Laevis (Janelia Research Campus, Howard Hughes Medical Institute, USA). We gratefully acknowledge the support of NVIDIA Corporation with the donation of the Titan Xp GPU used for this research. K.S.G. has received funding from the European Union's Horizon 2020 research and innovation program under the Marie Skłodowska-Curie Grant Agreement No. [750528]. M.L. thanks Prof. Stefan W. Hell for the research position in his department.

Supporting Information

# Spectral Cross-Cumulants for Multicolor Super-resolved SOFI Imaging

K. S. Grußmayer[1,2, #, *], S. Geissbuehler[2, #], A. Descloux [1,2], T. Lukes[1,2], M. Leutenegger[2,3], A. Radenovic[1], T. Lasser[2,4,*]

Affiliations

[1]École Polytechnique Fédérale de Lausanne, Laboratory of Nanoscale Biology, 1015 Lausanne, Switzerland

[2]École Polytechnique Fédérale de Lausanne, Laboratoire d'Optique Biomédicale, 1015 Lausanne, Switzerland

[3]Max-Planck Institute for Biophysical Chemistry, Department of NanoBiophotonics, Am Fassberg 11, 37077 Göttingen, Germany

[4]Max-Planck Institute for Polymer Research, Ackermannweg 10, 55128 Mainz, Germany

Contributions

[#]These authors contributed equally to this work.

Corresponding Authors:

Kristin Grußmayer, Theo Lasser

* email: kristin.grussmayer@epfl.ch, theo.lasser@epfl.ch

*Theory of spectral unmixing using spectral cross-cumulants*

*$n^{th}$ order spectral cross-cumulant analysis between two adjacent physical spectral channels*

For the case of 2 physical spectral channels (here: transmission channel T and reflection channel R, collecting fluorescence light in a specific wavelength range. A according to the combined spectral response of all the filters implemented in the microscope, with corresponding transmission $T_i$ and reflection coefficients $R_i$ per fluorophore species *i*) and a total of $N_c$ fluorophore species we write:

$$I_R(\boldsymbol{r}) = \sum_{i=1}^{N_c} R_i I_i(\boldsymbol{r})$$
$$I_T(\boldsymbol{r}) = \sum_{i=1}^{N_c} T_i I_i(\boldsymbol{r})$$
(S1)

with $I_i(\boldsymbol{r})$ being the intensity distribution of species *i* measured on detector pixel *r*. This linear system cannot be inverted to solve for the images of the isolated fluorophore species for more unknowns $I_1(\boldsymbol{r})$, $I_2(\boldsymbol{r})$, …, $I_{N_c}(\boldsymbol{r})$ than measurements, i.e. for $N_c > 2$ in this case.

However, if we assume stochastic, independent blinking of all the fluorescent emitters of the different species[1], we can apply cumulant analysis on the time series recorded in the transmission and reflection channels and generate an additional $n-1$ virtual channels by computing the $n^{th}$-order cross-cumulants (provided appropriate sampling of the PSFs). Due to the additivity property, the cumulant of multiple independent species corresponds to the sum of the cumulants of each individual species and we can rewrite:

$$\kappa_{n,(T,\ldots,R)}(\boldsymbol{r}) = \sum_{i=1}^{N_c} T_i^u R_i^{n-u} \kappa_n \{I_i(\boldsymbol{r},t)\}$$
(S2)

With $(T, \ldots, R)$ the set of *n* physical channels denoting the cross-cumulant that is computed using *u* pixels from the transmission channel and $n-u$ pixels from the reflection channel. $\kappa_n \{I_i(\boldsymbol{r},t)\} = \kappa_{n;i}$ denotes the $n^{th}$-order cumulant of the different fluorophore species *i*.

*General case of n^th-order spectral cross-cumulant analysis and $N_p$ physical color channels*

For the most general case of $N_p$ physical spectral channels (psc) and $N_c$ fluorescent species *i*, we can define a corresponding proportion $P_{psc,i}$ of the intensity that is directed into the specific spectral channel.

$$I_{psc}(\boldsymbol{r}) = \sum_{i=1}^{N_c} P_{psc,i} I_i(\boldsymbol{r}) \tag{S3}$$

If we assume stochastic, independent blinking of all the fluorescent emitters of the different species, we can again generate virtual channels by computing the $n^{th}$-order cross-cumulants. Due to the additivity, the cumulant of multiple independent species corresponds to the sum of the cumulants of each individual species and we can rewrite:

$$\kappa_{n,(psc_1,\ldots,psc_n)}(\boldsymbol{r}) = \sum_{i=1}^{N_c} \left( \prod_{j=1}^{n} P_{psc_j,i} \right) \kappa_n\{I_i(\boldsymbol{r},t)\} \tag{S4}$$

With $(psc_1, \ldots, psc_n)$ the set of physical spectral channels $psc_j \in \{1, \ldots, N_p\}$ denoting the cross-cumulant that is computed using pixels from the physical spectral channel $psc_j$. $\kappa_n\{I_i(\boldsymbol{r},t)\} = \kappa_{n;i}$ denotes the $n^{th}$-order cumulant of the different fluorophore species *i*. This computation of additional channels is the key to enable unmixing by inversion of the linear system of equations.

*Single-color single-species cumulant for m emitters:*

For *m* fluorescent emitters of a single fluorophore species *i* recorded in a single color channel, the $n^{th}$-order cumulant can be written as

$$\kappa_{n,i}(\boldsymbol{r}) \propto \varepsilon_i^n(\boldsymbol{r}) f_{n,i}(\rho_{on,i}; \boldsymbol{r}) \sum_{k=1}^{m} U_i^n(\boldsymbol{r} - \boldsymbol{r}_k), \tag{S5}$$

where $\varepsilon_i(r)$ is the the spatial distribution of the molecular brightness, $f_{n,i}(\rho_{on,i}; r)$ is the $n^{th}$-order cumulant of a Bernoulli distribution with on-time ratio $\rho_{on,i} = \frac{\tau_{on,i}}{\tau_{on,i}+\tau_{off,i}}$ and $U_i(r)$ is the system PSF for the fluorophore species $i^2$.

*Flattening:*

Using cross cumulants, virtual pixels are calculated in between the physical pixels acquired by the camera. Subsequently, proper weights are assigned to these virtual pixels in the so-called flattening operation assuming a known PSF[3] (see Equation S6) or optimal weights are calculated using a computationally demanding approach based on jackknife resampling[4].

$$\chi_{n;i}\left(r = \frac{1}{n}\sum_{k=1}^{n} r_k\right) = \frac{\kappa_{n;i}(r_1, \ldots, r_n)}{d_{n,k}(r_1, \ldots, r_n)} \tag{S6}$$

Where $d_{n,k}(r_1, \ldots, r_n) = \prod_{j<l}^{n} U_i\left(\frac{r_j - r_k}{\sqrt{n}}\right)$ is the distance factor.

In this study, we used the simple yet effective approach of weighing to the same mean within sub-grids of the image. Calculating the $n^{th}$ order cross-cumulant, we obtain $n-1$ virtual pixels in between each pair of pixels of the original pixel grid. For example, in the case of the 2$^{nd}$ order cumulant, we generate 3 virtual pixels for each physical pixel (i.e. there are 4 "pixel types" in the new, finer grid). This new grid can be divided into 4 mutually shifted sub-grids (each composed out of pixels of the same "pixel type"). These sub-grids represent the same image shifted by $p_s/n$, where p$_s$ is the projected pixel size of the original image and $n$ is the cumulant order. Assuming that these mutually shifted subsampled versions of the full image are supposed to have the same mean, the flattening can be performed by simply normalizing the sub-grids to the same mean value as the mean of the original image (i.e. sub-grid composed of the physical pixels).

*Transmission and reflection coefficients*

*From spectral data of the used fluorophores and filters*

The known fluorophore emission spectra are weighted with the spectral response curves of the reflection and transmission channels obtained from transmission data of the different (dichroic) filters that are implemented in the microscope (see *Microscope setup* section in Materials and methods of the main text). An example of the fluorophore Alexa Fluor 568 and the dichroic splitting ~ 594nm is provided in Figure S1.

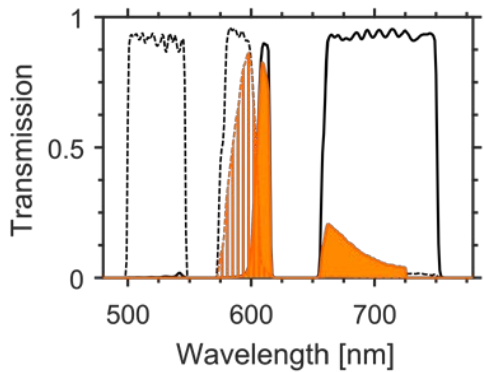

*Figure S1 Emission spectra of Alexa Fluor 568 multiplied with the spectral response curve of the reflection and transmission channel obtained by a dichroic color splitter (~594 nm) and a multi-band dichroic and emission filter to suppress the excitation laser light. Alexa Fluor 568 emission in the reflection channel (dark orange stripes), Alexa Fluor 568 emission in the transmission channel (dark orange), spectral response of the reflection channel (black dashed line) and of the transmission channel (black solid line).*

Here, we assume $D_T(\lambda) + D_R(\lambda) = 1$ and use the transmission data provided by the manufacturer (can also be measured in a spectrometer). Subsequently, the transmission $T_i$ and reflection coefficients $R_i$ per fluorophore species *i* can be calculated by integrating the transmitted (see Figure S1 dark orange) and reflected emission spectra (see Figure S1 dark orange stripes), respectively, and by normalizing with the total emission: $R, T = \frac{\int_0^\infty S_i(\lambda) D_{T,R}(\lambda) d\lambda}{\int_0^\infty S_i(\lambda) d\lambda} = \frac{I_{R,T}}{\sum(I_R + I_T)}$. The transmission coefficients for the fluorophore and filter combinations used in this work are provided in Table S1.

| Fluorophore | Dichroic mirror splitting fluorescence emission | Transmission coefficient T |
|---|---|---|
| *Alexa Fluor 488* | Laser Beamsplitter zt 594 RDC | *0.02* |
| *Alexa Fluor 488* | Beamsplitter HC BS 640 imaging | *0.04* |
| *Janelia Fluor 549* | Laser Beamsplitter zt 594 RDC | *0.16* |
| *Atto 565* | Laser Beamsplitter zt 594 RDC | *0.35* |
| *Atto 565* | Beamsplitter HC BS 640 imaging | *0.18* |
| *Abberior Flip 565* | Laser Beamsplitter zt 594 RDC | *0.26* |
| *Alexa Fluor 568* | Laser Beamsplitter zt 594 RDC | *0.47* |
| *Alexa Fluor 647* | Laser Beamsplitter zt 594 RDC | *0.98* |
| *Alexa Fluor 647* | Beamsplitter HC BS 640 imaging | *0.97* |

*Table S1 Transmission coefficients calculated according to the spectral response of the transmission channel for different fluorophores and dichroic mirrors in the emission path.*

*Experimental determination*

Cells are labelled with a single fluorophore species *i* and widefield images across the same filter combination as in the multicolour experiments are obtained in the reflection and emission channel. Subsequently, the transmission $T_i$ and reflection coefficients $R_i$ for fluorophore species *i* can be calculated by summing the background corrected transmitted and reflected intensity, respectively, and normalizing by the total emission.

***Optimization of multicolor SOFI via eigenvalue and eigenvector analysis***

The selection of the best combination of dyes and adequate filter sets can be an overwhelming and challenging task. In this section, we discuss a systematic approach to guide potential users in this process.

Each fluorophore is characterized by its emission spectrum $S_i(\lambda)$. If we assume that we use a perfect dichroic filter with a transmission function

$$D_T(\lambda; \lambda_D) = H(\lambda - \lambda_D)$$

where $H(\lambda)$ is the Heaviside step function and $\lambda_D$ is the characteristic wavelength of the dichroic, we can express the transmission and reflection coefficients $R_i, T_i$ as

$$T_i = \int_0^{\lambda_D} S_i(\lambda) d\lambda$$

$$R_i = \int_{\lambda_D}^{\infty} S_i(\lambda) d\lambda$$

where we assume that the spectrum $S_i(\lambda)$ is already normalized. Similarly, we can express the unmixing matrix $M$ as a function of $\lambda_D$

$$M(\lambda_D) = \begin{pmatrix} R_1^2 & R_2^2 & R_3^2 \\ T_1 R_1 & T_2 R_2 & T_3 R_3 \\ T_1^2 & T_2^2 & T_3^2 \end{pmatrix} = \begin{pmatrix} \left(\int_{\lambda_D}^{\infty} S_1(\lambda) d\lambda\right)^2 & \left(\int_{\lambda_D}^{\infty} S_2(\lambda) d\lambda\right)^2 & \dots \\ \left(\int_{\lambda_D}^{\infty} S_1(\lambda) d\lambda\right)\left(\int_0^{\lambda_D} S_1(\lambda) d\lambda\right) & \dots & \dots \\ \left(\int_0^{\lambda_D} S_1(\lambda) d\lambda\right)^2 & \dots & \dots \end{pmatrix}$$

The matrix $M(\lambda_D)$ is not invertible if one or more of its eigenvalues are equal to 0. In practice, the linear system is degraded by several noise sources and a matrix with an eigenvalue close to 0 is likely to be unstable. In order words, we need to optimizethe product $|\lambda_1||\lambda_2||\lambda_3|$ or maximize the smallest eigenvalue. Using $M(\lambda_D)$ we can compute the eigenvalues $\lambda_i$ for any set of fluorophores as a function of $\lambda_D$. Figure S2a shows the spectal response of three dyes Alexa Fluor 488, Atto 565, Alexa Fluor 647 and the filter set of the microscope. Figure S2b displays the absolute value of the three eigenvalues of $M(\lambda_D)$ as a function of $\lambda_D$ as well as their product. As expected, when $\lambda_D$ is smaller than 560nm or greater than 700nm, at least one eigenvalue tendsto 0, meaning that at least two dyes are completely reflected or transmitted. We observe a maximum for the eigenvalues product at around 600nm, which corresponds to the theoretical optimal splitting wavelength.

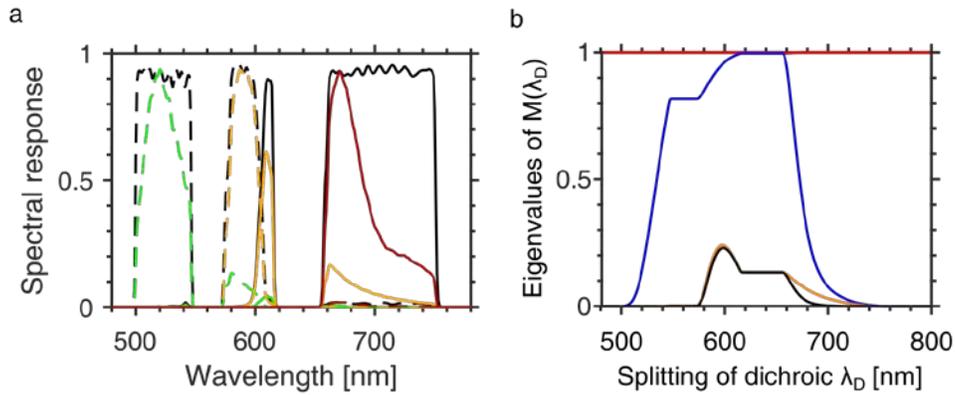

*Figure S2 Theoretical analysis of dyes-dichroic filter combinations. (a) Emission spectra of common organic fluorophores multiplied with the spectral response curve of the reflection and transmission channel obtained by a dichroic color splitter (~594 nm) and a multi-band dichroic and emission filter to suppress the excitation laser light. Alexa Fluor 488 (green), Atto565 (yellow), Alexa Fluor 647 (red), spectral response of the reflection channel (black dashed line) and of the transmission channel (black solid line). (b) Eigenvalues (red, blue and orange lines) of $M(\lambda_D)$ for $\lambda_D \in [450; 800nm]$ and product of the three eigenvalues (black line) with a maximum at 600nm indicating the theoretical best splitting ratio for this choice of dyes.*

In the case of a real dichroic with non-idealized reflection and transmission characteristics, we have to rewrite $R$ and $T$ as

$$T_{i,j} = \int_0^\infty S_i(\lambda) D_j(\lambda) d\lambda$$

$$R_{i,j} = \int_0^\infty S_i(\lambda) \left(1 - D_j(\lambda)\right) d\lambda$$

We can then rank any dyes and dichroic combination and select the $S_1(\lambda)$, $S_2(\lambda)$, $S_3(\lambda)$ and $D_j(\lambda)$ that produces the least singular matrix. In our case, the choice of ZT594RDC with a splitting at 594nm results in the eigenvalues: 1.05, 1.05 and 0.245 (product of 0.27), validating the choice of the dichroic for this specific set of dyes.

The generalization to more channels and dyes is straightforward and will just add additional vectorial components to this eigenvalue/vector analysis.

The unmixing matrix $M(\lambda_D)$ corresponding to the experiments and simulations in this work is diagonalized $M = Q \Lambda Q^{-1}$ with $Q = [v_1 \; v_2 \; v_3]$ and $\Lambda_{ii} = \lambda_i$, as the three eigenvectors form a basis. As $M(\lambda_D)$ needs to be invertible, we choose $\lambda_D$ such that $\lambda_i \neq 0$ (see discussion above). Thus, the dimension of the image of M is the same as the dimension of its domain and the rank of the unmixing matrix equals the number of color channels. This confirms that our cumulant analysis indeed provides an independent third channel allowing the unmixing of the three fluorophore species.

***Co-registration of physical color channels***

*Co-registration based on calibration measurements with beads*

An affine transformation and bilinear interpolation (Matlab) based on a calibration measurement with fluorescent beads that can be detected in both physical channels is applied to the transmission channel. We typically use ∅ 0.2 µm TetraSpeck beads or ∅ 0.17 µm orange beads from the PSF calibration kit (Invitrogen) dried on glass and covered in the supplied immersion medium.

*Co-registration based on image correlation*

The temporal standard deviation of the transmission and reflection channels is computed to generate two background free images. An affine transformation and bilinear interpolation (Matlab) based on the normalized cross-correlation between the two images is then applied to all the frames of the transmission channel.

*Simulations - multicolor SOFI with spectral unmixing*

To this end, we investigated the influence of different photophysical properties of the fluorophores on the performance of our multicolor analysis in simulations (for details on the simulations, please consult the main text and Materials and Methods). We simulate thin, densely labelled filaments that are partially overlapping, mimicking the cytoskeleton of cells. We first confirm multicolour imaging with fluorophores that range from green to (far infra-) red emission and verify that different photobleaching and blinking kinetics do not impair multicolour imaging as long as cumulant analysis is appropriately performed. Last, we show that our concept is also able to discriminate between three fluorophores with largely overlapping emission spectra.

We simulated filaments (dimensions ~ 5 µm × 0.04 µm) labelled with the commonly used fluorophores Alexa Fluor 488, Atto 565 and Alexa Fluor 647 (see Figure S2). We chose the photophysical parameters to resemble typical SOFI conditions (see Ref. [5] and Figure S3) of densely labelled structures with blinking, but multiple overlapping emitters. Multicolour analysis is performed using coregistration based on simulated calibration measurements with multicolour beads that appear 50:50 in both physical color channels. We can faithfully recover the appropriate color channels, as can be seen in Figure S4c-f. We estimated the residual crosstalk between the recovered channels by quantifying the remaining background corrected signal of the non-overlapping part of the filaments (see Table 2). The estimated crosstalk of Alexa Fluor 488 into the Atto 565 channel was highest and with only 5 % still very low. All other contributions from fluorophores to the "wrong" channels were smaller. We could not determine the residual crosstalk in experiments, as a sample with non-overlapping structures and the fluorophores in question was not available (note: a technical sample with entirely separate structures would be best). An extremely faint spot at the crossing of the Alexa Fluor 488 with the Alexa Fluor 647 filament is seen in the unmixed Atto 565 channel, most likely due to spurious correlation arising from the very dense blinking with a high on-ratio. For these densely labelled structures with high on-ratio blinking it suffices to analyse only a few hundred simulated frames (see Figure S5). This allows extremely fast and easy-to-implement multicolour imaging.

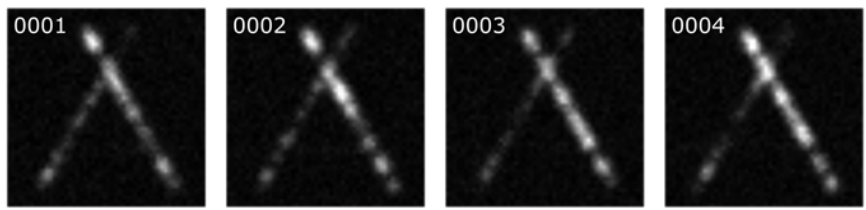

*Figure S3 Blinking of dense emitters with high on-ratio. Simulations of three fluorophores with emission in the green to (near infra-) red range. Alexa Fluor 488 (horizontal filament), Atto 565 (left, inclined to the right) and Alexa Fluor 647 (right, inclined to the left) at a density of 1000 fluorophores µm$^{-2}$ for a 48 pixel × 0.4 pixel filament (corresponding to 5.184 µm × ~43 nm), $I_{on}$ = 400 photons, on-ratio $\rho_{on}$ = 0.1 ($\tau_{on}$ = 20 ms, $\tau_{off}$ = 180 ms) and $\tau_{PB}$ = 80 s. Four raw frames from the transmission channel illustrate the simulated photophysics.*

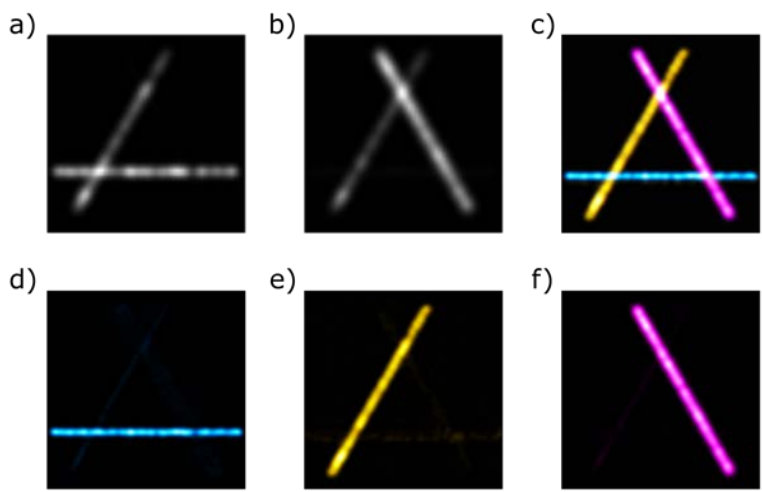

*Figure S4 Simulations of three fluorophores with emission in the green to (near infra-) red range. Alexa Fluor 488 (horizontal filament), Atto 565 (left, inclined to the right) and Alexa Fluor 647 (right, inclined to the left) at a density of 1000 fluorophores µm$^{-2}$ for a 48 pixel × 0.4 pixel filament (corresponding to 5.184 µm × ~43 nm), $I_{on}$ = 400 photons, on-ratio $\rho_{on}$ = 0.1 ($\tau_{on}$ = 20 ms, $\tau_{off}$ = 180 ms) and $\tau_{PB}$ = 80 s. a) and b) Average of 4000 frames in the reflection and transmission channel. c) Composite image of the balanced second-order SOFI images with d) Alexa Fluor 488 (cyan hot), e) Atto 565 (yellow hot) and f) Alexa Fluor 647 (magenta hot).*

| Fluorophore/channel | Alexa Fluor 488 | Atto 565 | Alexa Fluor 647 |
|---|---|---|---|
| **Alexa Fluor 488** | 100 | 5 | 1 |
| **Atto 565** | 2 | 100 | 2 |
| **Alexa Fluor 647** | 2 | 4 | 100 |

*Table 2 Relative crosstalk in % determined in the three-color simulated sample in Figure S4 with the fluorophores Alexa Fluor 488, Atto 565 and Alexa Fluor 647, when only background corrected pixels from regions without filament overlap were considered. The table is read line-wise with the fluorophore whose signal bleeds through listed on the left (e.g. read as: 2 % of Atto 565 bleeding into the unmixed Alexa Fluor 488 channel).*

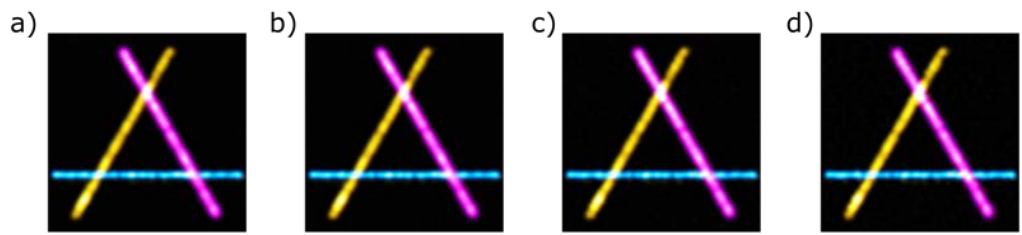

*Figure S5 Simulations of three fluorophores with varying number of frames used for analysis. Alexa Fluor 488 (horizontal filament), Atto 565 (left, inclined to the right) and Alexa Fluor 647 (right, inclined to the left) at a density of 1000 fluorophores µm$^{-2}$ for a 48 pixel × 0.4 pixel filament (corresponding to 5.184 µm × ~43 nm), $I_{on}$ = 400 photons, on-ratio $\rho_{on}$ = 0.1 ($\tau_{on}$ = 20 ms, $\tau_{off}$ = 180 ms) and $\tau_{PB}$ = 40 s. Composite image of the balanced second-order SOFI images with Alexa Fluor 488 (cyan hot), Atto 565 (yellow hot) and Alexa Fluor 647 (magenta hot) with a) 4000, b) 2000, c) 1000 and d) 500 frames used for analysis.*

*Fluorophores with different photobleaching kinetics*

To study the influence of differences in photobleaching, we considered Alexa Fluor 488, Atto 565 and Alexa Fluor 647 as above and only change the photobleaching time to 10, 40 and 80 s, respectively (see Figure S6). This already covers almost one order of magnitude difference in photostability. There is no noticeable change in the performance of our analysis, as the blinking kinetics and cumulant analysis are still appropriate for the fluorophore with the worst stability.

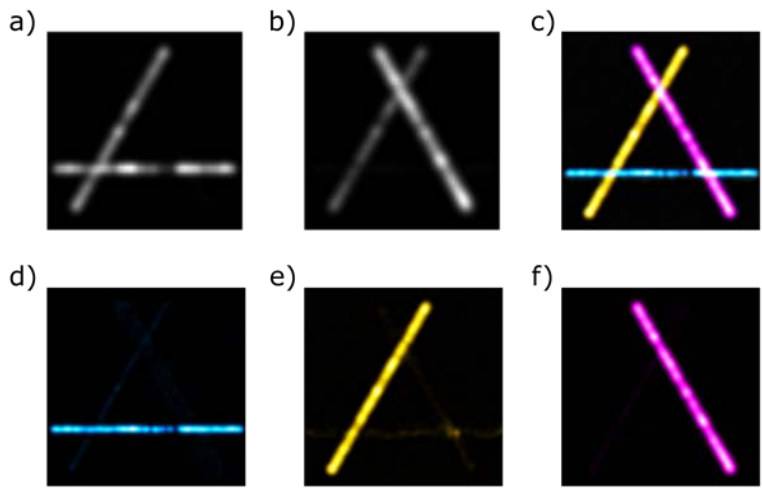

*Figure S6 Simulations of three fluorophores with different photobleaching kinetics. Alexa Fluor 488 (horizontal filament) with $\tau_{PB}$ = 10 s, Atto 565 (left, inclined to the right) with $\tau_{PB}$ = 40 s and Alexa Fluor 647 (right, inclined to the left) with $\tau_{PB}$ = 80 s at a density of ~1000 fluorophores µm$^{-2}$ for a 48 pixel × 0.4 pixel filament (corresponding to 5.184 µm × ~43 nm), $I_{on}$ = 400 photons, on-ratio $\rho_{on}$ = 0.1 ($\tau_{on}$ = 20 ms, $\tau_{off}$ = 180 ms). a) and b) Average of 4000 frames in the reflection and transmission channel. c) Composite image of the balanced second-order SOFI images with d) Alexa Fluor 488 (cyan hot), e) Atto565 (yellow hot) and f) Alexa Fluor 647 (magenta hot).*

*Fluorophores with different blinking kinetics*

Since not all fluorophores show the same blinking performance under identical experimental conditions[6], we tested the algorithm with different blinking off-times. As in the first simulations above, we considered Alexa Fluor 488, Atto 565 and Alexa Fluor 647 and only changed the on-ratio to 0.05, 0.1 and 0.01 ($\tau_{on}$ = 20 ms, $\tau_{off}$ = 380 ms, 180 ms and 1980 ms), respectively. This covers one order of magnitude difference in off-switching kinetics. The algorithm recovers all three color channels faithfully (see Figure S7).

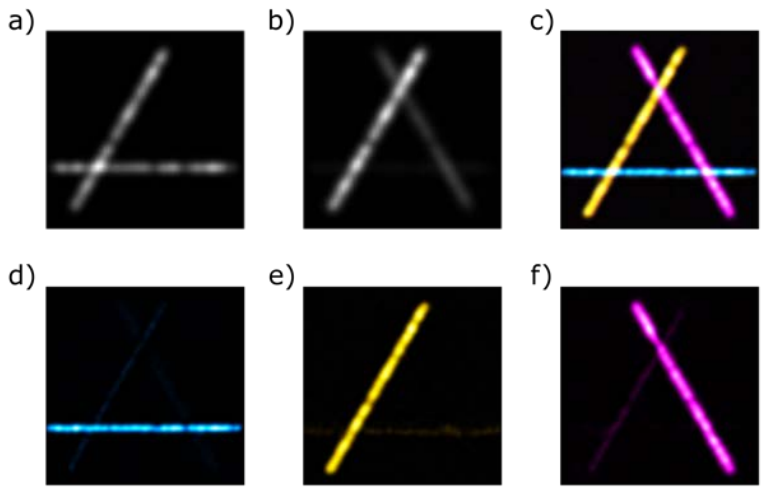

*Figure S7 Simulations of three fluorophores with different blinking kinetics. Alexa Fluor 488 (horizontal filament) with on-ratio $\rho_{on}$ = 0.05 ($\tau_{on}$ = 20 ms, $\tau_{off}$ = 380 ms), Atto 565 (left, inclined to the right) with on-ratio $\rho_{on}$ = 0.1 ($\tau_{on}$ = 20 ms, $\tau_{off}$ = 180 ms) and Alexa Fluor 647 (right, inclined to the left) with on-ratio $\rho_{on}$ = 0.01 ($\tau_{on}$ = 20 ms, $\tau_{off}$ = 1980 ms) at a density of ~1000 fluorophores µm$^{-2}$ for a 48 pixel × 0.4 pixel filament (corresponding to 5.184 µm × ~43 nm), $I_{on}$ = 400 photons and $\tau_{PB}$ = 80 s. a) and b) Average of 4000 frames in the reflection and transmission channel. c) Composite image of the balanced second-order SOFI images with d) Alexa Fluor 488 (cyan hot), e) Atto 565 (yellow hot) and f) Alexa Fluor 647 (magenta hot).*

*Fluorophores with different brightness*

Similarly, it is difficult to achieve equal brightness for all fluorophores in experiments due to different spectral properties and blinking behaviour. As in the first simulations above, we considered Alexa Fluor 488, Atto 565 and Alexa Fluor 647 and only changed the intensity from $I_{on}$ = 400, 600 to 800 photons ($\tau_{on}$ = 20 ms, $\tau_{off}$ = 180 ms), respectively. The algorithm recovers all three color channels faithfully (see Figure S8) and the reconstructed filaments have no obvious difference in contrast.

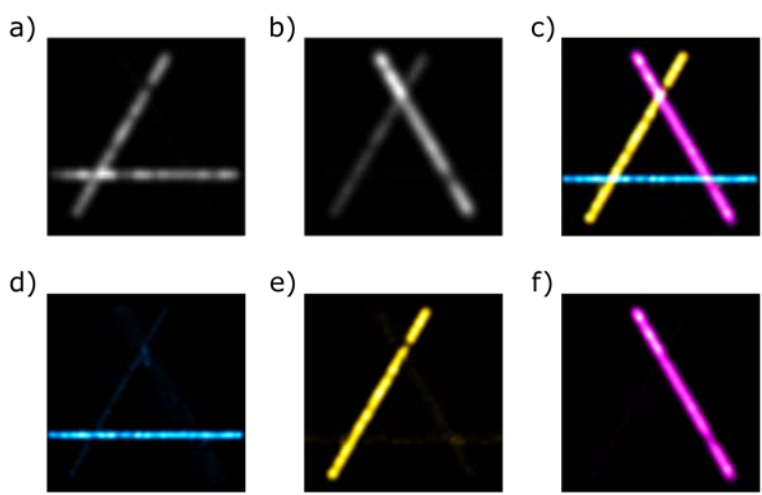

*Figure S8 Simulations of three fluorophores with different intensity. Alexa Fluor 488 (horizontal filament) $I_{on}$ = 400 photons, Atto 565 (left, inclined to the right) with $I_{on}$ = 600 photons and Alexa Fluor 647 (right, inclined to the left) with $I_{on}$ = 800 photons ($\tau_{on}$ = 20 ms, $\tau_{off}$ = 180 ms) at a density of ~1000 fluorophores µm$^{-2}$ for a 48 pixel × 0.4 pixel filament (corresponding to 5.184 µm × ~43 nm) and $\tau_{PB}$ = 80 s. a) and b) Average of 4000 frames in the reflection and transmission channel. c) Composite image of the balanced second-order SOFI images with d) Alexa Fluor 488 (cyan hot), e) Atto 565 (yellow hot) and f) Alexa Fluor 647 (magenta hot).*

*Fluorophores with largely overlapping spectra*

Next, we changed only the spectral parameters and simulated three fluorophores with largely overlapping emission spectra separated by about 10 nm only, such that the fluorophores can all be excited by one laser line (here: e.g. 561nm for Abberior Flip 565 ($\lambda_{\frac{abs}{em},max}$ = 566/ 580 nm), Atto 565 ($\lambda_{\frac{abs}{em},max}$ = 564/ 590 nm) and Alexa Fluor 568 ($\lambda_{\frac{abs}{em},max}$ = 578/ 603 nm), see Figure S9). The multicolour results in Figure S10 show that our algorithm can even separate these fluorophores that are impossible to distinguish from the average diffraction limited reflection and transmission images. A detailed inspection of the images shows faint ghost images of filaments from the other channels. We again calculated the residual crosstalk between the recovered channels by quantifying the remaining background corrected signal of the non-overlapping part of the filaments (see Table 3). The estimated crosstalk of Alexa Fluor 568 and Abberior Flip 565 into the Atto 565 channel was highest with 14%. All other contributions from fluorophores to the "wrong" channels were smaller.

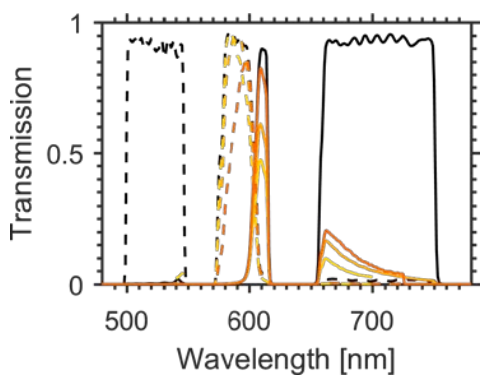

*Figure S9 Emission spectra of spectrally closely overlapping common organic fluorophores multiplied with the spectral response curve of the reflection and transmission channel obtained by a dichroic color splitter (~594 nm) and a multi-band dichroic and emission filter to suppress the excitation laser light. AbberiorFlip 565 (yellow), Atto 565 (orange), Alexa Fluor 568 (dark orange), spectral response of the reflection channel (black dashed line) and of the transmission channel (black solid line).*

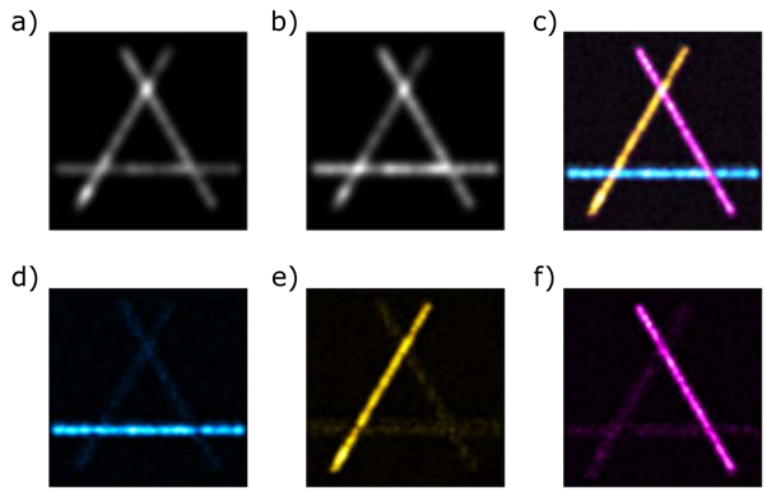

*Figure S10 Simulations of three fluorophores with spectrally very similar emission in the yellow to red range. Alexa Fluor 568 (horizontal filament), AbberiorFlip 565 (left, inclined to the right) and Atto 565 (right, inclined to the left) at a density of ~1000 fluorophores µm$^{-2}$ for a 48 pixel × 0.4 pixel filament (corresponding to 5.184 µm × ~43 nm), $I_{on}$ = 400 photons, on-ratio $\rho_{on}$ = 0.1 ($\tau_{on}$ = 20 ms, $\tau_{off}$ = 180 ms) and $\tau_{PB}$ = 80 s. Channels were overlaid as simulated. a) and b) Average of 4000 frames in the reflection and transmission channel. c) RGB composite image of the unmixed and deconvolved second-order SOFI images with d) Alexa Fluor 568 (cyan hot), e) AbberiorFlip 565 (yellow hot) and f) Atto 565 (magenta hot).*

| Fluorophore/channel | Alexa Fluor 568 | Abberior Flip 565 | Atto 565 |
|---|---|---|---|
| **Alexa Fluor 568** | 100 | 12 | 14 |
| **Abberior Flip 565** | 6 | 100 | 14 |
| **Atto 565** | 9 | 12 | 100 |

*Table 3 Relative crosstalk in % determined in the three colour simulated sample in Figure S10 with the fluorophores Alexa Fluor 568, Abberior Flip 565 and Atto 565, when only background corrected pixels from regions without filament overlap were considered. The table is read line-wise with the fluorophore whose signal bleeds through listed on the left (e.g. read as: 6 % of Abberior Flip 565 is bleeding into the unmixed Alexa Fluor 568 channel).*

*Multicolor SOFI with spectral unmixing: additional data*

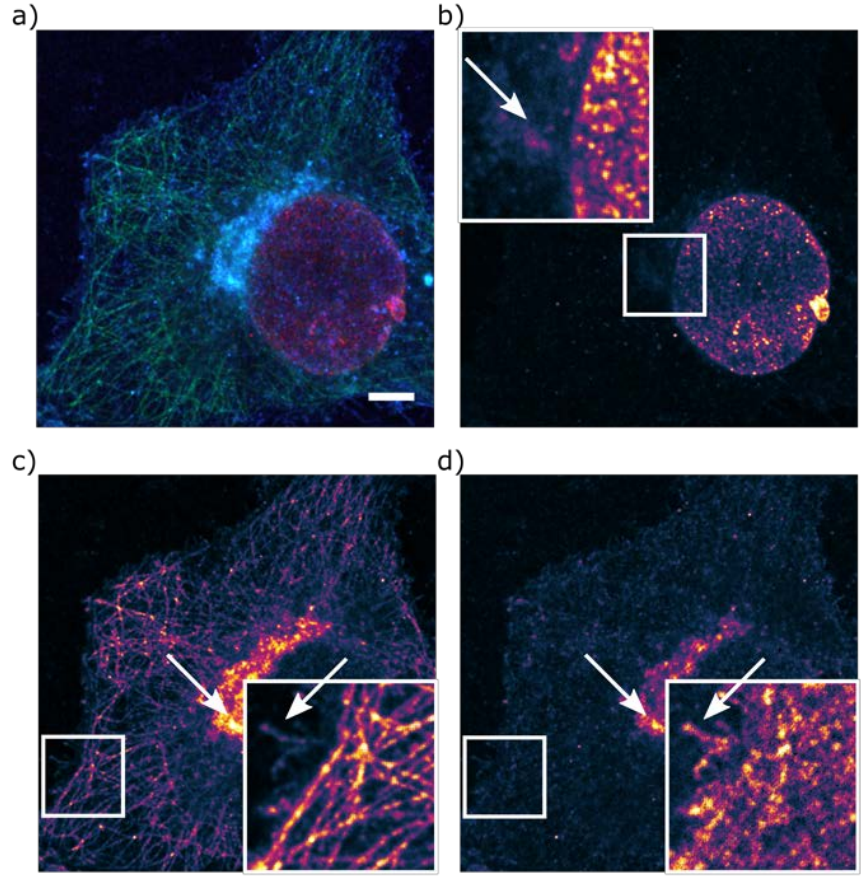

*Figure S11 Second order spectral cross-cumulant images of the cytoskeleton, nucleus and cellular membranes of HeLa cells. a) RGB composite image of the second order spectral cross-cumulant images with b) $\kappa_{2,TT}$(red), c) $\kappa_{2,RR}$(green) and $\kappa_{2,RT}$(blue). The separate cross-cumulant images are displayed using the morgenstemning colormap[7]. Scale bar 5 µm. Data from Figure 3. 200mM MEA with oxygen scavenging and about 0.5 kWcm$^{-2}$ 488nm, 1.25 kWcm$^{-2}$ 561nm and 1.3 kWcm$^{-2}$ 635nm illumination intensity. The arrows illustrate crosstalk of the fluorophores in the three spectral cross-cumulant channels.*

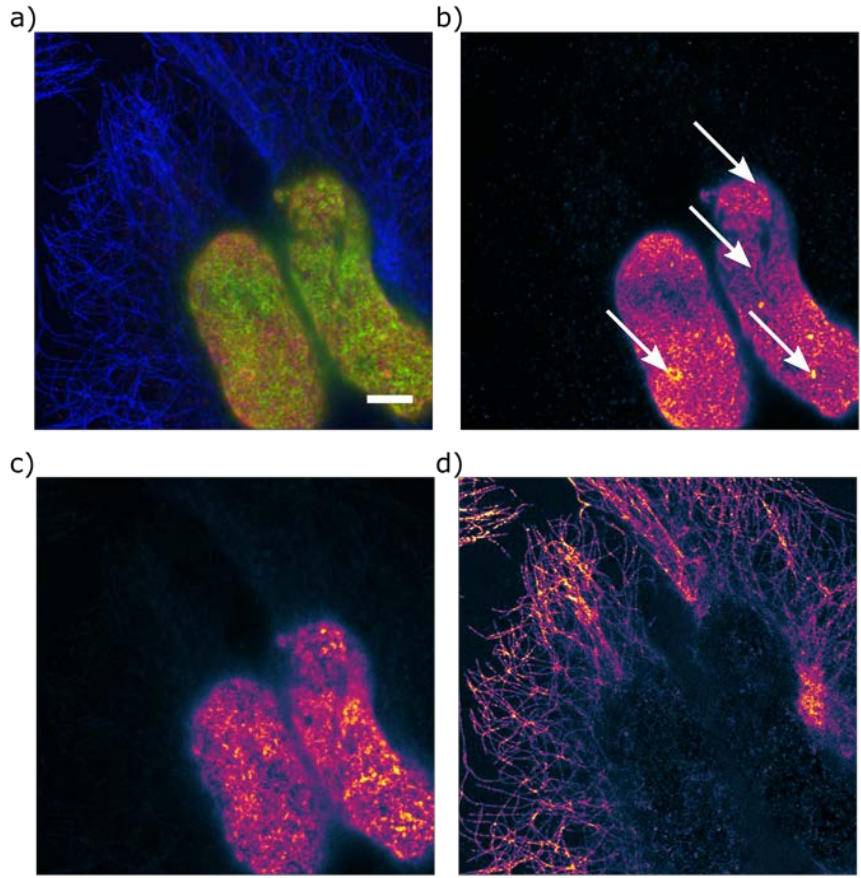

*Figure S12 Multicolor SOFI of the cytoskeleton and nucleus of HeLa cells. a) RGB composite image of the unmixed, flattened and deconvolved second order SOFI images with b) Alexa Fluor 647 secondary antibody stained nuclear membrane (red), c) Hoechst-Janelia Fluor 549 DNA labeling (green) and. d) Alexa Fluor 488 secondary antibody stained microtubules (blue). The separate unmixed images are displayed using the morgenstemning colormap[7]. The arrows in b) indicate typical features of Lamin B staining such as folds in the nuclear membrane. Scale bar 5 µm. Data from Figure 4. 50mM MEA with oxygen scavenging and about 0.5 kWcm$^{-2}$ 488nm, 1.25 kWcm$^{-2}$ 561nm and 1.3 kWcm$^{-2}$ 635nm illumination intensity.*

## Multicolor SOFI with spectral unmixing: additional experiments

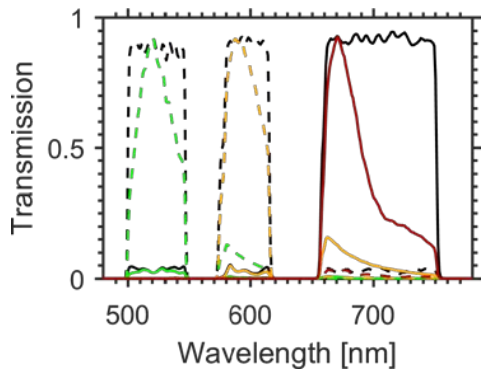

Figure S13 Emission spectra of common organic fluorophores multiplied with the spectral response curve of the reflection and transmission channel obtained by a dichroic color splitter (~640 nm) and a multi-band dichroic and emission filter to suppress the excitation laser light. Alexa Fluor 488 (green), Atto 565 (orange), Alexa Fluor 647 (red), spectral response of the reflection channel (black dashed line) and of the transmission channel (black solid line).

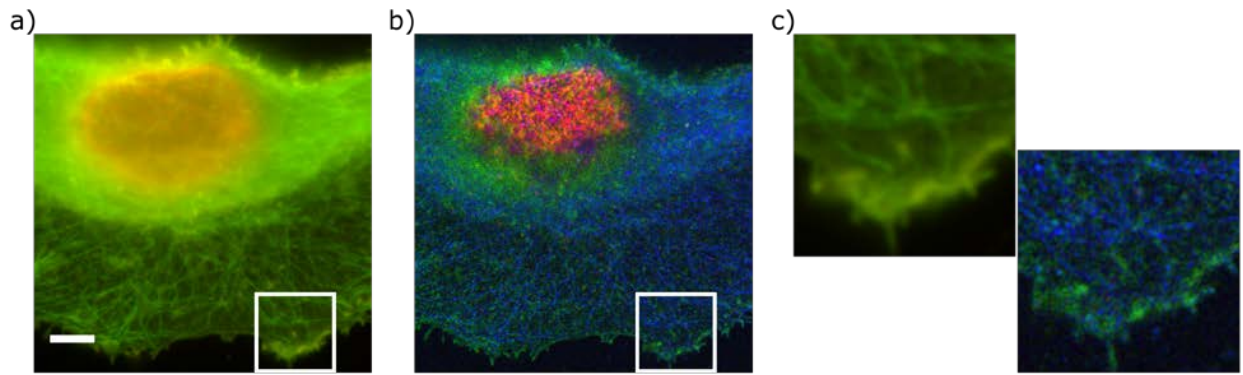

Figure S14 Multicolor SOFI of the cytoskeleton, nucleus and cellular membranes of HeLa cells with a dichroic splitting ~640 nm. a) Overlay of the average of the time series acquired in the reflection (green) and transmission channel (red) using 200mM MEA with oxygen scavenging and about 0.5 kWcm$^{-2}$ 488nm, 1.25 kWcm$^{-2}$ 561nm and 0.85 kWcm$^{-2}$ 635nm illumination intensity. b) RGB composite image of the unmixed and deconvolved second order SOFI images with Alexa Fluor 488 secondary antibody stained microtubules (blue), wheat germ agglutinin-Atto565 labeling (green) and Alexa Fluor 647 secondary antibody stained nuclear membrane (red). c) Close-up of the ROI indicated in a) and b), respectively. Scale bar 5 µm.

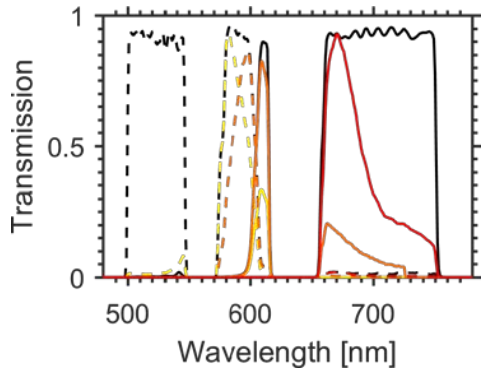

*Figure S15 Emission spectra of common organic fluorophores multiplied with the spectral response curve of the reflection and transmission channel obtained by a dichroic color splitter (~594 nm) and a multi-band dichroic and emission filter to suppress the excitation laser light. Janelia Fluor 549 (yellow), Alexa Fluor 568 (dark orange), Alexa Fluor 647 (red), spectral response of the reflection channel (black dashed line) and of the transmission channel (black solid line).*

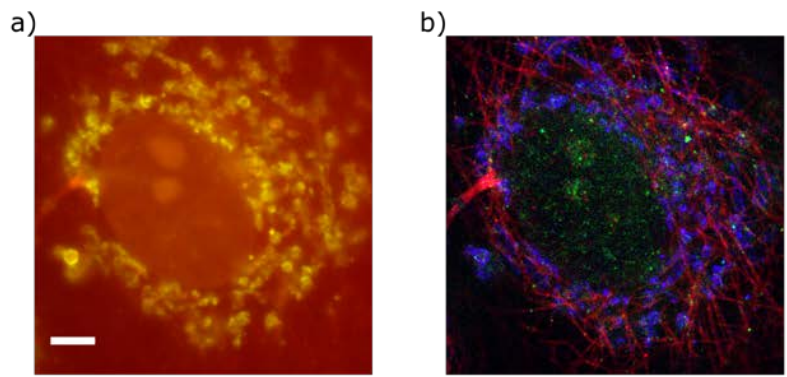

*Figure S16 Multicolor SOFI of the cytoskeleton, nucleus and mitochondria of COS-7 cells with a dichroic splitting ~594 nm. a) Overlay of the average of the time series acquired in the reflection (green) and transmission channel (red) using 50mM MEA with oxygen scavenging and about 1.25 kWcm$^{-2}$ 561nm and 1.3 kWcm$^{-2}$ 635nm illumination intensity. b) RGB composite image of the unmixed and deconvolved second order SOFI images with Alexa Fluor 568 secondary antibody stained mitochondria (blue), Hoechst-Janelia Fluor 549 labeling (green) and Alexa Fluor 647 secondary antibody stained microtubules (red). Scale bar 5 μm. The staining of the DNA is sparse and was performed with 5 times lower concentration as for HeLa cells shown in the main text.*